\documentclass[runningheads]{llncs}
\usepackage{graphicx}
\usepackage{amsmath}
\usepackage{amssymb}
\usepackage{booktabs}
\usepackage{multirow}
\usepackage{xcolor}
\usepackage{makecell}
\usepackage{enumitem}
\usepackage{graphicx}
\usepackage{hyperref} 
\usepackage{multirow}
\usepackage{graphicx}
\usepackage{amsmath}
\usepackage{amssymb}
\usepackage{booktabs}
\usepackage{multirow}
\usepackage{xcolor}
\usepackage{makecell}
\usepackage{enumitem}
\begin{document}
\title{On-the-Fly Test-time Adaptation for Medical Image Segmentation}
\author{Anonymous MICCAI Submission} 

\author{Jeya Maria Jose Valanarasu, Pengfei Guo, Vibashan VS, and Vishal M. Patel } 

\institute{Johns Hopkins University}
%
%
\authorrunning{JMJ Valanarasu et al.}
%
%
\maketitle              
\begin{abstract}
One major problem in deep learning-based solutions for medical imaging is the drop in performance when a model is tested on a data distribution different from the one that it is trained on.  Adapting the source model to target data distribution at test-time is an efficient solution for the data-shift problem. Previous methods solve this by adapting the model to target distribution by using techniques like entropy minimization or regularization. In these methods, the models are still updated by back-propagation using an unsupervised loss on complete test data distribution. In real-world clinical settings, it makes more sense to adapt a model to a new test image on-the-fly and avoid model update during inference due to privacy concerns and lack of computing resource at deployment. To this end, we propose a new setting - On-the-Fly Adaptation which is zero-shot and episodic (\emph{i.e.}, the model is adapted to a single image at a time and also does not perform any back-propagation during test-time). To achieve this, we propose a new framework called Adaptive UNet where each convolutional block is equipped with an adaptive batch normalization layer to adapt the features with respect to a domain code. The domain code is generated using a pre-trained encoder trained on a large corpus of medical images. During test-time, the model takes in just the new test image and generates a domain code to adapt the features of source model according to the test data. We validate the performance on both 2D and 3D data distribution shifts where we get a better performance compared to previous test-time adaptation methods. Code is available at \href{https://github.com/jeya-maria-jose/On-The-Fly-Adaptation}{https://github.com/jeya-maria-jose/On-The-Fly-Adaptation}

\end{abstract}

\section{Introduction}

Image segmentation is a major task in medical imaging as it is essential for computer-aided diagnosis and image-guided surgery systems. In the past few years, deep learning-based solutions have been widely popular for medical image segmentation. Many convolutional methods \cite{ronneberger2015u,zhou2018unet++,milletari2016v,islam2018ischemic,valanarasu2020kiu,valanarasu2021kiu}  and transformer-based methods \cite{chen2021transunet,jose2021medical} have been proposed for various medical image segmentation tasks showing very good performance. However, a major problem with deep neural networks (DNN) is that they are highly dependent on the dataset that they are trained on. If a DNN is trained on a specific dataset and tested on a different dataset, the performance usually drops  even if they are of the same modality. This happens due to occurrence of many shifts like camera/scanner parameters, resolution, intensity, and contrast variations. This drop in performance makes DNN-based solutions for medical imaging tasks impractical to be adopted for real-time clinical use. For clinical use, the model needs to be robust as there can be small changes in the test data distribution even if they are of the same modality. 

Many domain adaptation techniques for medical image segmentation have looked into solving this problem \cite{guan2021domain}. However, this setting assumes that we have access to the source model, source data as well as the target data. Another setting very close to real-time is fully test-time adaptation \cite{wang2021tent} where we assume that we do not have access to the  source data and adapt the model to the target data by performing one back propagation per sample. However, the model is adapted to the complete test distribution as the model weights are  updated for at least one complete epoch. This setting can be considered one-shot adaptation as the model sees all the data in the distribution at least once. This can also be extended for few-shot adaptation to further adapt the model during test time. However, this setting is also not clinically deployable as we need a complete distribution to perform the adaptation and get the new model weights. Also, performing back-propagation during test-time means that we still have to do some training during the deployment-time although it is unsupervised. 

In this work, we propose a clinically motivated setting called On-the-Fly test-time adaptation where the model adapts to a single image/volume at a time. Here, we do not perform any back-propagation during test time and just attempt to adapt our  model to the new data instance. Also, as the model is reset for every data instance there is no need to assume the availability of the complete target distribution to perform adaptation since patient data may come with privacy concerns. On-the-Fly Adaptation is a more useful scenario in the current trend as there has been a shift of laboratory to bed-side settings for medical imaging \cite{vashist2017point}.  To solve this problem, we propose a new framework called Adaptive-UNet where the model is equipped with adaptive batch-norm layers in both the encoder and decoder to adapt to a select domain code. 

In summary, the following are the contributions of this work: 1) We introduce On-the-Fly Adaptation which is more closer to real-world clinical scenarios where the adaptation is zero-shot and episodic removing the assumption of the availability of complete target distribution and back-propagation during test-phase. 2) We propose Adaptive-UNet, a new framework that learns to adapt to a new test data instance making use of a domain code and adaptive batch normalization. 3) We validate our method for 9 domain shifts in medical image segmentation for 2D fundus images and 3D MRI volumes where we get better performance than recent test-time  adaptation methods.

\section{Related Works}

\noindent \textbf{Unsupervised Domain Adaptation} for medical image segmentation is a widely explored topic. Methods like feature alignment using adversarial training \cite{javanmardi2018domain,panfilov2019improving}, disentangling the representation \cite{yang2019unsupervised}, ensembling and using soft labels \cite{perone2019unsupervised} have been proposed. These methods, however, use the training distribution of both source and target data for adaptation which is not always feasible for medical imaging due to privacy concerns.

\noindent \textbf{Source-free Unsupervised Domain Adaptation} works for medical image segmentation  assume no availability of source data. In \cite{bateson2020source}, a label-free entropy loss is defined over target distribution with a domain-invariant prior. In \cite{chen2021source}, an uncertainty aware denoised pseudo label method is proposed.

\noindent \textbf{Test-time Adaptation} methods such as TENT \cite{wang2021tent} uses entropy minimization of batch norm statistics to adapt to a new target distribution. Recently, Hu et al. \cite{hu2021fully} proposed using new losses like regional nuclear norm and contour regularization to improve test-time performance for medical image segmentation. Self domain adapted networks \cite{he2020self} use auto-encoder based adaptors to rapidly adapt to a new task at test-time. Karani et al. \cite{karani2021test} proposed a per-test-image adaptation method where they adapt the image so as to obtain plausible segmentation. They still update the weights during test time by assessing the similarity of a given segmentation to those in the source data.

\section{On-the-Fly Adaptation}
\vspace{-1.5 em}
\begin{figure}[htbp]
	\centering
	\includegraphics[width=1\linewidth]{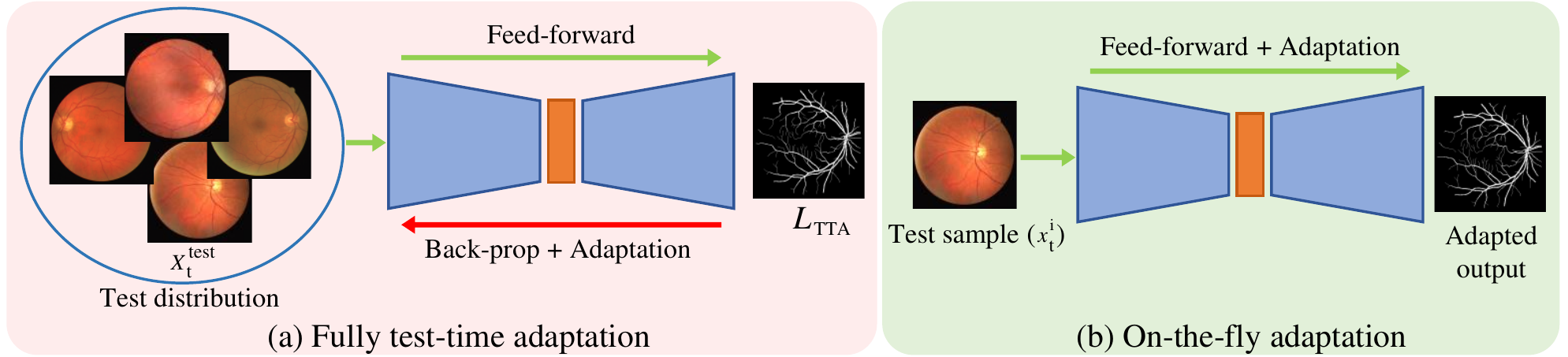}
	
	\vskip-11pt\caption{Comparison between (a) Test-time adaptation, and (b) On-the-Fly Adaptation. }  
	\label{intro}
\end{figure}

 Let $X$ and $Y$ represent the set of input data and labels, respectively. Let us assume that the source distribution is represented as $X_s$, $Y_s$ and the target distribution as  $X_t$, $Y_t$. In normal source training, we train the model using $X_s^{train}$, $Y_s^{train}$ and test the model on $X_s^{test}$. For direct testing (no adaptation), we use the model trained on $X_s^{train}$, $Y_s^{train}$ and test on $X_t^{test}$. The best performance would be obtained when the model is trained using $X_t^{train}$, $Y_t^{train}$ and tested on  $X_t^{test}$ which corresponds to the oracle performance. 

In general test-time adaptation, we assume the  availability of the entire target test distribution $X_t^{test}$ during test-time. The weights are optimized on the gradients calculated using an unsupervised loss function loss $\mathcal{L}_{tta}(X^{test}_t)$ obtained using test data distribution $X_t^{test}$.  Some works like \cite{karani2021test} perform adaptation for each test-image. However, they do optimize the weights according to the test-image at hand during inference. Optimizing the network weights across all the data in test distribution and later using the new model to validate on $X_t^{test}$ again is not suitable in a clinical setting. It makes more sense to adapt the model for each test-image as using the entire test distribution in medical setting involves using a variety of patient data at test-time. It is also difficult to update model weights during deployment time as it requires massive computational power.

In the proposed On-the-Fly adaptation, we focus on adapting to a single test image/volume at a time as illustrated in Fig. \ref{intro}. Instead of assuming we have the entire target test distribution $X_t^{test}$, we assume we only have a single data-instance $x_t^i$ which is actually the case during clinical deployment. This makes On-the-Fly adaptation episodic as it resets to original weight for adapting to each data-instance. Also, we constrain the setting to not perform any back-propagation during the test-phase as it involves the availability of compute power or some cloud resource during testing. This makes On-the-Fly adaptation zero-shot as it does not really involve any gradient back-propagation during test-time. The setting of On-the-Fly adaptation is summarized in Table 1 and compared against other frameworks.

\begin{table*}[t!]
\centering
\caption{
Comparison between different adaptation problem setups.
Notation: source ($s$), target ($t$), data distribution $X$, label distribution $Y$.
}
\label{tab:settings}

\resizebox{1\linewidth}{!}{
    \begin{tabular}{c|c|c|c|c}
    \toprule
    \bf Setting & \bf Source data & \bf Target data & \bf Train loss & \bf Test loss \\ \hline
    
    Source training & $X^{train}_s, Y^{train}_s$ & - & $\mathcal{L}_{det}(X^{train}_s, Y^{train}_s)$ & - \\
    Oracle & - & $X^{train}_t, Y^{train}_t$ & $\mathcal{L}_{det}(X^{train}_t, y^{train}_t)$ & - \\
    Unsupervised domain adaptation & $X^{train}_s$, $Y^{train}_s$ & $X^{train}_t$ & $\mathcal{L}_{det}(X^{train}_s, Y^{train}_s)$ + $\mathcal{L}_{da}(X^{train}_s, X^{train}_t)$ & - \\
    Source free adaptation  & - & $X^{train}_t$ & $\mathcal{L}_{da}(X^{train}_t)$ & - \\
    Test-time training  & $X^{train}_s$, $Y^{train}_s$ & $X^{test}_t$ & $\mathcal{L}_{det}(X^{train}_s, Y^{train}_s)$ + $\mathcal{L}_{aux}(X^{train}_s)$ & $\mathcal{L}_{aux}(X^{test}_t)$ \\
    Fully test-time adaptation & - & $X^{test}_t$ & - & $\mathcal{L}_{tta}(X^{test}_t)$ \\
    On-the-Fly adaptation & - & $x^{i}_t \epsilon X^{test}_t$  & - & - \\
    \bottomrule
    \end{tabular}
}

\end{table*}

\section{Method - Adaptive UNet}

To solve On-the-Fly Adaptation, we propose an Adaptive UNet framework where we make use of adaptive batch normalization and a domain prior. 

\begin{figure}[htbp]
	\centering
	\includegraphics[width=1\linewidth]{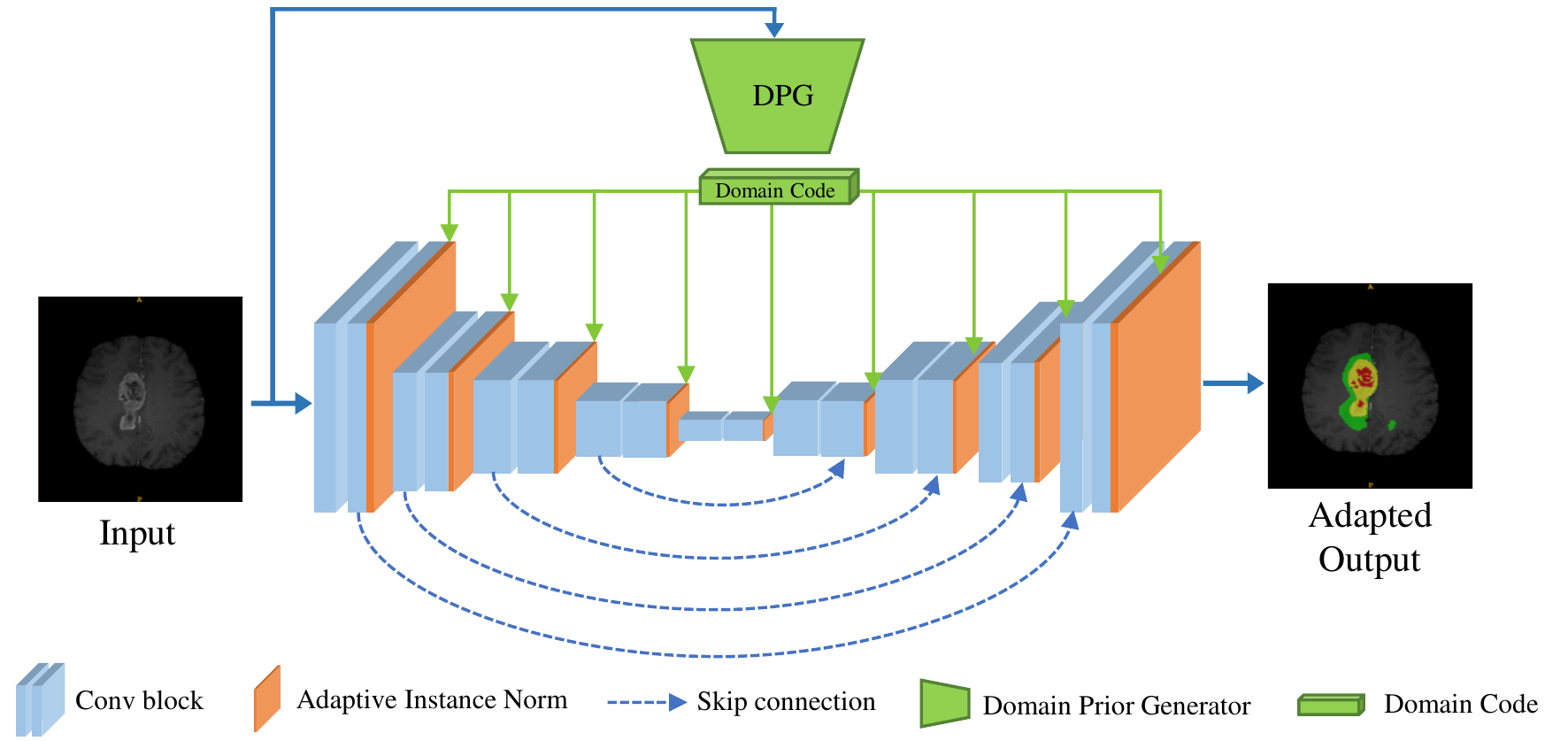}
	
	\vskip-11pt\caption{Overview of Adaptive UNet framework. }  
	\label{arch}

\end{figure}

\noindent \textbf{Network Details:} We follow the skeleton of a generic UNet architecture \cite{ronneberger2015u}. We use 5 conv blocks in both the encoder and decoder, respectively. Each conv block in the encoder consists of a conv layer, adaptive batch normalization, ReLU activation and a max-pooling layer. Each conv block in the decoder consists of a conv layer, adaptive batch normalization, ReLU activation and an upsampling layer. For upsampling, we use bilinear interpolation. For our experiments on 3D volumes, we use a 3D UNet architecture \cite{cciccek20163d} with the same setup replacing 2D conv layer with 3D conv layers, 2D max-pooling with 3D max-pooling and bilinear upsampling with trilinear upsampling.

\noindent \textbf{Adaptive Batch Normalization:} Batch Normalization (BN) layers \cite{ioffe2015batch} are used in DNNs to mitigate the issue of internal co-variate shifts. It normalizes the features in the network helping in training and faster convergence. BN can be defined as:
\setlength{\belowdisplayskip}{0pt} \setlength{\belowdisplayshortskip}{0pt}
\setlength{\abovedisplayskip}{0pt} \setlength{\abovedisplayshortskip}{0pt}
\begin{equation}
    z = \gamma \left(\frac{X - \mu(X)}{\sigma(X)}\right) + \beta,
\end{equation}
where $x$ is the input batch, $z$ represents the output, $\mu$ represents the mean $E[X]$, $\sigma$ represents standard deviation  $\sqrt{Var(X)}$. Here, $\gamma$ and $\beta$ are learnable parameters  which control the scaling and shifting while normalizing.

Adaptive Instance normalization (AdaIN) \cite{huang2017arbitrary} is used to align the mean and standard deviation of two feature codes (usually one being context and another being style). AdaIN can be defined as:
\begin{equation}
    z = \sigma(y) \left(\frac{x - \mu(x)}{\sigma(x)}\right) + \mu(y),
\end{equation}
where $x$ and $y$ are the two feature codes and $z$ is the normalized output.

In Adaptive UNet, we make use of Adaptive Batch Normalization (AdaBN) which basically learns scaling and shifting operation while adaptively normalizing the batch statistics between two codes. AdaBN can be defined as:
\begin{equation}
     z = \gamma \left(\sigma(Y) \left(\frac{X - \mu(X)}{\sigma(X)}\right) + \mu(Y)\right) + \beta.
\end{equation}
Note that our formulation is a bit different from AdaBN as explained in \cite{li2016revisiting} as it tries to shift the model to test data's mean and standard deviation instead of aligning them. Here, we align the codes while also learning how to align them by learning the shift and scale parameters. In Adaptive UNet, the input to AdaBN layers are the feature codes from the UNet represented as $X$ and domain codes $Y$ generated from the domain prior generator.

\noindent \textbf{Domain Prior Generator:} The Domain Prior Generator (DPG) is an encoder of a pre-trained auto-encoder. We first pre-train a UNet as an auto-encoder for medical images. This task is self-supervised as we just try to predict the original image while feeding an augmented version of the data as input. Doing this helps the model learn an abstract code in the latent space. We train the model on a variety of medical data consisting of different modalities. More details can be found in the supplementary file. We make sure that the distribution of data that we conduct experiments to validate Adaptive UNet do not overlap with the data that the auto-encoder is trained on. However, it does have the  data of similar modality. This helps the encoder generate different domain codes for different modalities. For example, two images of T1 MRI would have their corresponding domain codes closer in the latent space when compared to T1 MRI and T2 MRI.

\noindent \textbf{Training source model:} During the training phase of source model, we feed in the input image to both the encoder of UNet and the pre-trained domain prior generator. The domain code obtained from the domain prior generator is passed to the AdaBN layers in the encoder and decoder. The feature maps are normalized according to the domain code using AdaBN. So, in the training itself the model has learned to adapt to the domain code of the current modality/distribution. The learnable parameters $\gamma$ and $\beta$ of AdaBN layers learn the scale and shift necessary to adapt to the style code at each level to provide the optimal segmentation prediction. Note that the weights are updated only for the UNet segmentation network. The pre-trained domain prior generator is frozen during training the source model.

\noindent \textbf{Inference on target data instance:} When a model trained on $X_s^{train}$ is validated on a target data instance $x_t$, we pass the image $x_t$ to both the domain prior generator and the source model. First, we generate the new domain code using the domain prior generator for the new image $x_t$. Next we pass this domain code to all the AdaBN layers in Adaptive UNet. During feed forward, the features extracted at each layer of Adaptive UNet are adapted to the new domain code. So, the model is thus adapted according to the code of the new modality/target domain. There is no back-propagation involved as the features are adapted in feed-forward itself. Also, as the model weights are not changed, this framework is episodic and does not depend on the entire test data distribution for validation. An overview of the framework is illustrated in Fig. \ref{arch}.

\section{Experiments and Results}

\noindent \textbf{Datasets:} For 2D experiments, we focus on the task of retinal vessel segmentation from fundus images. We make use of the following datasets: CHASE \cite{fraz2012ensemble}, RITE \cite{hu2013automated} and HRF \cite{odstrvcilik2009improvement}. CHASE contains 28 retina images with a resolution of 999$\times$960 collected from 14 school children with a hand-held Nidek NM-200-D fundus camera. RITE consists of 40 images of resolution 768$\times$584 collected from people aging from 25 to 90 using a  Canon CR5 non-mydriatic 3CCD camera. HRF contains 18 images collected from 18 human subjects using a Canon CR-1 fundus camera of around resolution 3504$\times$2336. There exists a domain shift among these datasets as they vary with respect to camera properties, age of patients and resolution etc. The datasets are separated into a randomized 80-20 split wherever test split was not given.  For 2D experiments, DPG is pre-trained on fundus images from \cite{fun}

For 3D experiments, we focus on brain tumor segmentation from MRI volumes. We make use of the BraTS 2019 dataset \cite{menze2014multimodal,bakas2017advancing,bakas2018identifying} which consists of four modalities- FLAIR, T1, T1ce and T2. We study the domain shift problems between these four modalities for volumetric segmentation of brain tumor. This is a multi-class segmentation problem with 4 labels. We randomly split the dataset into 266 for training and 69 for validation. We do this as the ground truth is not provided publicly for the original validation dataset. For the MRI experiments, we pre-train DPG on MRI images from Kaggle MRI dataset \cite{mri} and IXI dataset \cite{mri2}.

\noindent \textbf{Implementation Details:} We use Pytorch framework for implementing Adaptive UNet. For 2D experiments, we use a combination of binary cross entropy (BCE) and dice loss to train Adaptive UNet. The loss $\mathcal{L}$ between the prediction $\hat{y}$ and the target $y$ is formulated as:
\begin{equation}
    \mathcal{L} = \lambda  BCE(\hat{y}, y) + Dice(\hat{y}, y).
\end{equation}
We use an Adam optimizer with a learning rate of 0.0001 and momentum of 0.9. We also use a cosine annealing learning rate scheduler with a minimum learning rate upto 0.00001. The batch size is set equal to 8. For 3D experiments, we use a similar loss but use a learning rate of 0.001 while also reducing the batch size to 2. More details can be found in code and the supplementary file.

\noindent \textbf{Performance Comparison:} We compare our proposed method with recent test-time adaptation methods like TENT \cite{wang2021tent}, Hu et al. \cite{hu2021fully} (RN+CR loss) , and self domain adapted network (SDA) \cite{he2020self}. In Table \ref{quan1}, we present the results of 2D experiments for 6 different domain shifts in fundus image. In Table \ref{quan2}, we present the results of 3D experiments for 3 different domain shifts in MRI modality. In both the tables, the first row corresponds to the direct adaptation results where we train the model on source domain and report the results of those models while testing on the target domain without any adaptation. The last row corresponds to the oracle which is the maximum possible performance when the model is trained on the target train distribution and tested on the target test distribution. Note that the 3D experiments have two target-training configurations- Uni-modal and Multi-modal. Uni-modal oracle corresponds to the configuration where we only use one modality and multi-modal oracle corresponds to the case where we use all four modalities to train the model. The compared test-time adaptation methods are presented in both one-shot and ten-shot settings. In one-shot setting, the model weights are adapted by back-propagation for one epoch using the test distribution. In ten-shot setting, the model weights are adapted for ten epochs by back-propagation using the test distribution. For our proposed method, we adapt the model once per image during feed-forward.

\begin{table}[htbp]
\centering

\caption{Results for 2D Domain shifts. Numbers correspond to dice score. \textbf{\textcolor{red}{Red}} and \textbf{\textcolor{blue}{Blue}} corresponds to the first and second best performing methods respectively. }

\resizebox{0.9\columnwidth}{!}{%
\begin{tabular}{c|c|c|c|c|c|c|c}
\Xhline{3\arrayrulewidth}
Type                                  & Method           & \begin{tabular}[c]{@{}c@{}}CHASE \\ -\textgreater \\ HRF\end{tabular} & \begin{tabular}[c]{@{}c@{}}CHASE \\ -\textgreater \\ RITE\end{tabular} & \begin{tabular}[c]{@{}c@{}}HRF \\ -\textgreater \\ CHASE\end{tabular} & \begin{tabular}[c]{@{}c@{}}HRF \\ -\textgreater \\ RITE\end{tabular} & \begin{tabular}[c]{@{}c@{}}RITE \\ -\textgreater \\ CHASE\end{tabular} & \begin{tabular}[c]{@{}c@{}}RITE \\ -\textgreater \\ HRF\end{tabular} \\ \Xhline{3\arrayrulewidth}
Source-Training                       & Direct Testing & \textbf{\textcolor{blue}{61.95}}                                                                  & 70.88                                                                  & 31.65                                                                 & \textbf{\textcolor{red}{62.87}}                                                                & 44.65                                                                  & 59.67                                                                \\ \hline
\multirow{3}{*}{\begin{tabular}[c]{@{}c@{}}Test-Time Adaptation\\ (One-shot)\end{tabular}} & SDA \cite{he2020self}               & 11.19                                                                 & 6.22                                                                   & 3.64                                                                  & 4.75                                                                 & 0.91                                                                   & 5.15                                                                 \\
                                      & TENT \cite{wang2021tent}              & 61.20                                                                  & \textbf{\textcolor{blue}{73.65 }}                                                                  & 19.20                                                                  & 58.64                                                                & 63.76                                                                  & 57.14                                                                \\
                                      & RN+CR \cite{hu2021fully}            & 61.31                                                                 & 72.85                                                                  & 21.20                                                                & 57.35                                                                & \textbf{\textcolor{blue}{63.95 }}                                                                  & 59.13                                                                \\ \hline
                                      \multirow{3}{*}{\begin{tabular}[c]{@{}c@{}}Test-Time Adaptation\\ (Ten-shot)\end{tabular}} & SDA \cite{he2020self}               & 3.11                                                                 & 5.13                                                                   & 1.84                                                                  & 5.48                                                                 & 7.33                                                                   & 5.15                                                                 \\
                                      & TENT \cite{wang2021tent}              & 61.17                                                                  & 73.62                                                                  & 19.22                                                                  & 58.64                                                                & 63.72                                                                  & 57.10                                                                \\
                                      & RN+CR  \cite{hu2021fully}           & 61.42                                                                 & 72.88                                                                  & \textbf{\textcolor{blue}{21.22}}                                                                  & 57.32                                                                & 63.91                                                                  & \textbf{\textcolor{blue}{59.15}}                                                                \\ \hline
\begin{tabular}[c]{@{}c@{}}\textbf{On-the-Fly Adaptation}\\ \textbf{(Zero-shot)}\end{tabular}                & \begin{tabular}[c]{@{}c@{}}\textbf{Adaptive UNet}\\ \textbf{(Ours)}\end{tabular}     & \textbf{\textcolor{red}{70.19}}                                                                 & \textbf{\textcolor{red}{74.27}}                                                                  & \textbf{\textcolor{red}{50.27}}                                                                 & \textbf{\textcolor{blue}{59.59}}                                                                & \textbf{\textcolor{red}{65.98}}                                                                  & \textbf{\textcolor{red}{63.14}}                                                                \\ \Xhline{3\arrayrulewidth}

Target-Training                       & Oracle            & 76.88                                                                 & 77.69                                                                  & 75.78                                                                 & 77.69                                                                & 75.78                                                                  & 76.88                                                                \\ \Xhline{3\arrayrulewidth}
\end{tabular}

}

\label{quan1}
\end{table}

From Tables 2 and 3, it can be inferred that there is a considerable drop in performance while directly testing the source model on the target domain. This drop is expected as there exits a domain shift between the source and the target distribution. SDA does not perform well, especially in relatively small datasets (Table~2). Since SDA requires training a set of auto-encoders to provide supervision during test-time adaption, only training on a small amount of data may result in overfitting and consequently lower adaptation performance. TENT and RN+CR methods improve the performance in most cases as they try to reduce the entropy and regularize the batch-norm statistics for the target distribution. Our proposed method shows a considerable improvement over the direct testing as well as test-time baselines on almost all domain shifts achieving state-of-the-art adaptation performance.  Note that brain tumor segmentation from 3D volumes is a multi-class segmentation problem which shows that our method can be successfully adopted for multi-class problems as well.

\begin{figure}[htbp]
	\centering
	\includegraphics[width=1\linewidth]{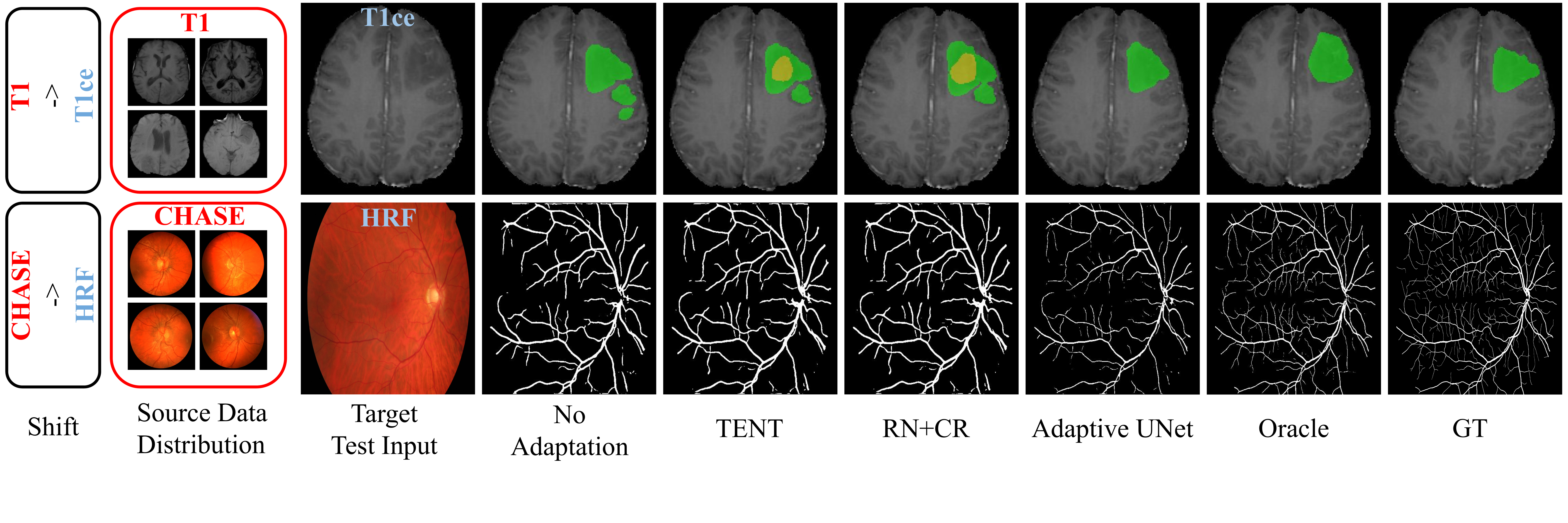}
	
	\vskip-25pt\caption{Qualitative Results.}  
	\label{qual}

\end{figure}

We also provide sample qualitative results in Fig. \ref{qual}. It can be observed that without any adaptation, the segmentation predictions are noisy and contain over-segmentation. While recent test-time methods improve the prediction, they still suffer from mis-classification of pixels (see MRI predictions in Fig. \ref{qual} for TENT/RN+CR) and also over-segmentation (see fundus predictions in Fig. \ref{qual} for TENT/RN+CR). Our method achieves  good segmentation prediction that is very close to the oracle prediction and the ground-truth.

\begin{table}[htbp]
\centering
\caption{Results for 3D Domain shifts. Numbers correspond to dice score reported in the following order:  WT/TC/ET. WT = Whole Tumor, TC = Tumor Core, ET = Enhancing Tumor. \textbf{\textcolor{red}{Red}} and \textbf{\textcolor{blue}{Blue}} corresponds to the first and second best performing methods respectively.}

\resizebox{1\columnwidth}{!}{%
\begin{tabular}{c|c|ccc}
\Xhline{3\arrayrulewidth}
Type                                                                                       & Method                                                         & \multicolumn{1}{c|}{T1 --\textgreater T1ce} & \multicolumn{1}{c|}{T1ce --\textgreater T1} & FLAIR --\textgreater T1ce \\ \hline
Source-Training                                                                            & Direct Testing                                              & \multicolumn{1}{c|}{48.74/52.35/36.48}  & \multicolumn{1}{c|}{59.25/33.78/9.02}   & \textbf{\textcolor{blue}{24.35}}/39.51/\textbf{\textcolor{red}{29.38}}     \\ \hline
\multirow{3}{*}{\begin{tabular}[c]{@{}c@{}}Test-Time Adaptation\\ (One-shot)\end{tabular}} & SDA \cite{he2020self}                                                           & \multicolumn{1}{c|}{10.89/48.98/29.47}  & \multicolumn{1}{c|}{12.03/6.52/2.25}    & 13.37/16.72/7.62      \\
                                                                                           & TENT \cite{wang2021tent}                                                          & \multicolumn{1}{c|}{57.41/55.31/39.79}  & \multicolumn{1}{c|}{\textbf{\textcolor{red}{68.25}}/46.99/6.61}   & 24.33/\textbf{\textcolor{blue}{41.95}}/22.90     \\
                                                                                           & RN+CR  \cite{hu2021fully}                                                        & \multicolumn{1}{c|}{55.21/54.62/38.21}  & \multicolumn{1}{c|}{\textbf{\textcolor{blue}{67.56}}/47.21/5.52}   & 24.31/41.85/22.82     \\ \hline
                                                                                           \multirow{3}{*}{\begin{tabular}[c]{@{}c@{}}Test-Time Adaptation\\ (Ten-shot)\end{tabular}} & SDA \cite{he2020self}                                                           & \multicolumn{1}{c|}{14.21/52.65/32.34}  & \multicolumn{1}{c|}{8.91/2.96/10.82}    & 14.01/28.11/13.88      \\
                                                                                           & TENT \cite{wang2021tent}                                                          & \multicolumn{1}{c|}{\textbf{\textcolor{blue}{55.47}}/52.89/39.17}  & \multicolumn{1}{c|}{66.23/42.52/\textbf{\textcolor{red}{12.14}}}   & 19.96/36.25/22.50     \\
                                                                                           & RN+CR  \cite{hu2021fully}                                                        & \multicolumn{1}{c|}{57.30/\textbf{\textcolor{blue}{54.68}}/\textbf{\textcolor{blue}{40.01}}}  & \multicolumn{1}{c|}{60.82/\textbf{\textcolor{blue}{42.87}}/\textbf{\textcolor{blue}{12.11}}}   & 19.39/33.96/17.09    \\ \hline
\begin{tabular}[c]{@{}c@{}}\textbf{On-the-Fly Adaptation}\\ \textbf{(Zero-shot)}\end{tabular}                & \begin{tabular}[c]{@{}c@{}}\textbf{Adaptive UNet}\\ \textbf{(Ours)}\end{tabular} & \multicolumn{1}{c|}{\textbf{\textcolor{red}{60.66}}/\textbf{\textcolor{red}{58.73}}/\textbf{\textcolor{blue}{39.30}}}  & \multicolumn{1}{c|}{65.08/\textbf{\textcolor{red}{48.09}}/8.78}   & \textbf{\textcolor{red}{24.86}}/\textbf{\textcolor{red}{42.27}}/\textbf{\textcolor{blue}{29.20}}     \\ \Xhline{3\arrayrulewidth}
\multirow{2}{*}{Target-Training}                                                           & Uni-Modal Oracle                                            & \multicolumn{1}{c|}{73.49/74.54/67.97}  & \multicolumn{1}{c|}{70.88/56.60/26.70}  & 73.49/74.54/67.97     \\ \cline{2-5}
                                                                                           & Multi-Modal Oracle                                             & \multicolumn{3}{c}{91.06/70.09/78.97}                                                                             \\ \Xhline{3\arrayrulewidth}
\end{tabular}
}

\label{quan2}
\end{table}

\noindent \textbf{Discussion: } From our experiments, we find that our method works pretty well for 2D experiments when compared to 3D experiments.  This observation can be understood as the 2D experiments consider domain shifts within the same modality with differences in camera/sensor properties and type of patient. The 3D experiments consider cross-modality domain shifts where the MRI sequences are themselves different. This is a more difficult task as each sequence extracts different types of features. However, we get a considerable boost over other test-time methods even though our setting is episodic and zero-shot. 


\section{Conclusion}

In this work, we propose a new adaptation setting called On-the-Fly adaptation. In this setting, we constrain the adaptation to be episodic and zero-shot thus assuming no availability of the entire target distribution and model update during test-time. This makes On-the-Fly adaptation very close to the scenario while deploying deep-learning models in real-world clinical settings. We propose a new framework- Adaptive UNet to solve this adaptation problem by making using of adaptive batch normalization and domain priors. We validate our model on both 2D and 3D domain shifts of fundus images and MRI volumes and show that the proposed method achieves a competitive performance over the recent test-time adaptation methods even with the tighter constraint of On-the-Fly adaptation.

\bibliographystyle{splncs04}
\bibliography{tta}

\end{document}